\documentclass[a4paper,fleqn,usenatbib]{mnras}
\usepackage{graphicx}
\usepackage{rotating}
\usepackage{lscape}
\usepackage{morefloats}
\usepackage{amssymb}
\DeclareMathAlphabet{\mathpzc}{OT1}{pzc}{m}{it}

\def\lsim{\,\lower2truept\hbox{${<\atop\hbox{\raise4truept\hbox{$\sim$}}}$}\,}
\def\gsim{\,\lower2truept\hbox{${> \atop\hbox{\raise4truept\hbox{$\sim$}}}$}\,}

\title[ALMA calibrators]{ALMA photometry of extragalactic radio sources}
\author[M. Bonato et al.]
{M. Bonato$^{1,2}$\thanks{matteo.bonato@inaf.it},
E. Liuzzo$^{1}$,
D. Herranz$^{3}$,
J. Gonz\'alez-Nuevo$^{4}$,
L. Bonavera$^{4}$,
M. Tucci$^{5}$,
\newauthor
M. Massardi$^{1}$,
G. De Zotti$^{2}$,
M. Negrello$^{6}$
and M. A. Zwaan$^{7}$\\
$^{1}$INAF$-$Istituto di Radioastronomia, and Italian ALMA Regional Centre, Via Gobetti 101, I-40129, Bologna, Italy\\
$^{2}$INAF$-$Osservatorio Astronomico di Padova, Vicolo Osservatorio 5, I-35122, Padova, Italy \\
$^{3}$Instituto de F\'isica de Cantabria (CSIC-Universidad de Cantabria),
Avda. de los Castros s/n, Santander, Spain \\
$^{4}$Departamento de F\'isica, Universidad de Oviedo, C. Federico Garc\'ia Lorca 18, E-33007 Oviedo, Spain\\
$^{5}$D\'epartement de Physique Th\'eorique and Center for Astroparticle Physics, Universit\'e de Gen\`eve\\
$^{6}$School of Physics and Astronomy, Cardiff University, The Parade, Cardiff CF24 3AA, UK\\
$^{7}$European Southern Observatory, Karl-Schwarzschild-Str. 2, D-85748 Garching, Germany}

\pagerange{\pageref{firstpage}--\pageref{lastpage}} \pubyear{2019}

\def\LaTeX{L\kern-.36em\raise.3ex\hbox{a}\kern-.15em
    T\kern-.1667em\lower.7ex\hbox{E}\kern-.125emX}
\def\simlt{\mathrel{\rlap{\lower 3pt\hbox{$\sim$}}\raise 2.0pt\hbox{$<$}}}
\def\simgt{\mathrel{\rlap{\lower 3pt\hbox{$\sim$}}\raise 2.0pt\hbox{$>$}}}

\begin{document}


\maketitle

\begin{abstract}
We present a new catalogue of ALMA observations of 3,364 bright, compact
radio sources, mostly blazars, used as calibrators.  These sources were
observed between May 2011 and July 2018, for a total of 47,115 pointings in
different bands and epochs. We have exploited the ALMA data to validate the
photometry given in the new \textit{Planck} Multi-frequency Catalogue of
Non-thermal sources (PCNT), for which an external validation was not possible
so far. We have also assessed the positional accuracy of \textit{Planck}
catalogues and the PCNT completeness limits, finding them to be consistent
with those of the Second \textit{Planck} Catalogue of Compact Sources. The
ALMA continuum spectra have allowed us to extrapolate the observed radio
source counts at 100 GHz to the effective frequencies of ALMA bands 4, 6, 7,
8 and 9 (145, 233, 285, 467 and 673\,GHz, respectively), where direct
measurements are scanty, especially at the 3 highest frequencies. The results
agree with the predictions of the Tucci et al. model C2Ex, while the
model C2Co is disfavoured.
\end{abstract}

\begin{keywords}
Radio sources: statistics -- Galaxies: active -- AGN: radio continuum --
submillimetre: galaxies
\end{keywords}

\section{Introduction}\label{sect:introduction}

Although a substantial progress on the characterization of millimeter and
sub-millimeter properties of extragalactic radio sources has been made in
recent years mainly thanks to surveys with the Wilkinson Microwave Anisotropy
Probe \citep[WMAP;][]{Bennett2013}, the \textit{Planck} satellite \citep{PCCS2,
PCNT}, the South Pole Telescope \citep[SPT;][]{Mocanu2013} and the Atacama
Cosmology Telescope \citep[ACT;][]{Marsden2014}, the available information is
still limited.

However, an unprecedented amount of multi-frequency and multi-epoch photometric
data on radio sources in this spectral region is being provided by the Atacama
Large Millimeter/submillimeter Array (ALMA). This is because ALMA uses
mm/sub-mm bright compact radio sources as calibrators to fix the flux density
scale, to determine the bandpass response, and to calibrate amplitude and phase
of the visibilities of the science targets. Observations of calibrator sources
(mainly point-like sources) are made for every science project. Each calibrator
is generally observed several times, in connection with different targets, on
different days, in various bands and array configurations.

\citet{Bonato2018} have published a catalogue of ALMA flux density measurements
of 754 calibrators for a total of 16,263 observations in different bands and
epochs (ALMACAL catalogue). These flux densities were uniformly
measured from a collection of ALMA images, thus obtaining robust measurements
for both resolved and non-resolved sources. The calibration and imaging
analyses are described in \citet{Bonato2018}. Almost all sources ($\sim$97\%)
were found to be blazars.

Combining such catalogue with the ``ALMA Calibrator Source Catalogue''
(ACSC)\footnote{Available at \url{https://almascience.eso.org/sc/}}, we have
collected ALMA observations for 3,364 bright, (mostly) compact radio sources.
These sources were observed between May 2011 and July 2018, for a total of
47,115 observations in different bands and epochs. We have 25,907 observations
of 3,310 different sources in band 3 (84$-$116\,GHz), 671 observations of 171
sources in band 4 (125$-$163\,GHz), 8,467 observations of 885 sources in band 6
(211$-$275\,GHz), 11,415 observations of 2,201 sources in band 7
(275$-$373\,GHz), 394 observations of 88 sources in band 8 (385$-$500\,GHz),
253 observations of 59 sources in band 9 (602$-$720\,GHz) and 8 observations of
6 sources in band 10 (787$-$950\,GHz).

The combined catalogue, referred to as the ALMA Calibrator Catalogue (ACC), contains the ALMA name of the source, its classification, redshift (when
available), equatorial coordinates (J2000), flux density measurements with
their uncertainties, effective observing frequency, date and time of
observations. One example of the information provided is in
Table~\ref{tab:fluxes}.

A search of the literature has yielded redshifts for 2245 sources (67\%). About
41\% (1391) of our sources are listed in the 5th edition of the Roma
Multi-frequency Catalogue of
Blazars\footnote{\url{http://www.ssdc.asi.it/bzcat/}}
\citep[BZCAT;][]{Massaro2009} where they are classified into 5 sub-classes:
FSRQs, BL Lacs, BL Lacs-galaxy dominated, Blazars of uncertain type, BL Lac
candidates. We have classified the remaining objects following
\citet{Bonato2018}.

The uncertainties on ALMA flux densities are essentially given by the
calibration uncertainty, whose value is still being debated within the
ALMA community. An accurate calibration is difficult to achieve due to
variability of the emissive and absorptive properties of the Earth's atmosphere
and to the lack of astronomical sources that could serve as accurate flux
standards. Following \citet{Bonato2018} we adopt at 5\% calibration uncertainty
for ALMACAL sources. This value was found to be consistent with the median
absolute differences among measurements in Bands 3 and 6 within short
time-spans (30\,d in the source frame), expected to be only weakly affected by
variability \citep[see Sect.~4 of][]{Bonato2018}.

We have made a similar check on ACSC sources. We selected observations within 30\,d in the source frame. For
this time span we found median absolute differences of 6\% for Bands 3 and 6,
and of 8\% for Band 7. There are no multiple measurements within this time span
in Bands 4, 8 and 9, so that no estimates could be obtained. These somewhat
larger uncertainties for ACSC compared to ALMACAL sources may be due to the
fact that the flux densities of the latter sources were uniformly measured from
a collection of ALMA images. The measured differences are however consistent
with the uncertainties given in the ACSC, taking into account that outliers are
to be expected since blazars show variability also on short timescales. We have
therefore adopted the ACSC uncertainties. 


\begin{table*}
\caption{Example of the catalogue content. The full catalogue is available online and on the
website of the Italian ALMA Regional Center (ARC; \url{http://arc.ia2.inaf.it})}.
\centering
\begin{tabular}{lccccccccc}
\hline
ALMA name & Class.$^{1}$ & z & RA [deg] & DEC [deg] & Flux density [Jy] & Uncert. [Jy] & band & $\nu$ [GHz] & Date of obs.$^{2}$\\
\hline
J2148+0657 & 1 & 0.89 & 327.0227 & 6.9607 & 2.0900 & 0.1000 & 3 & 98.21 & 2012-06-30T00:00:00 \\
 & & & & & 	1.9400 & 0.1000 & 3 & 109.74 & 2012-06-30T00:00:00 \\
 & & & & & 	1.9900 & 0.1000 & 3 & 98.21 & 2012-07-29T00:00:00 \\
 & & & & & 	1.8400 & 0.1100 & 3 & 109.74 & 2012-07-29T00:00:00 \\
 & & & & & 	1.9500 & 0.2200 & 3 & 106.25 & 2012-08-04T00:00:00 \\
 & & & & & 	2.0900 & 0.2100 & 3 & 94.35 & 2012-08-04T00:00:00 \\
 & & & & & 	1.9900 & 0.0500 & 3 & 98.21 & 2012-08-09T00:00:00 \\
 & & & & & 	1.8600 & 0.0500 & 3 & 109.74 & 2012-08-09T00:00:00 \\
 & & & & & 	1.9700 & 0.0600 & 3 & 98.21 & 2012-08-31T00:00:00 \\
 & & & & & 	1.8500 & 0.0700 & 3 & 109.74 & 2012-08-31T00:00:00 \\
 & & & & & 	1.9500 & 0.0400 & 3 & 98.21 & 2012-09-01T00:00:00 \\
 & & & & & 	1.8400 & 0.0500 & 3 & 109.74 & 2012-09-01T00:00:00 \\
 & & & & & 	1.8500 & 0.1100 & 3 & 106.25 & 2012-09-02T00:00:00 \\
 & & & & & 	1.9900 & 0.1300 & 3 & 94.35 & 2012-09-02T00:00:00 \\
 & & & & & 	1.9000 & 0.1400 & 3 & 106.25 & 2012-09-02T00:00:00 \\
 & & & & & 	2.0300 & 0.1400 & 3 & 94.35 & 2012-09-02T00:00:00 \\
 & & & & & 	1.8800 & 0.0900 & 3 & 99.2 & 2012-10-06T00:00:00 \\
 & & & & & 	1.7300 & 0.0900 & 3 & 108.76 & 2012-10-06T00:00:00 \\
 & & & & & 	1.0100 & 0.0500 & 6 & 221.0 & 2012-10-06T00:00:00 \\
 & & & & & 	0.6600 & 0.0300 & 7 & 343.25 & 2012-10-06T00:00:00 \\
 & & & & & 	1.9100 & 0.0300 & 3 & 98.21 & 2012-10-18T00:00:00 \\
 & & & & & 	1.7800 & 0.0500 & 3 & 109.74 & 2012-10-18T00:00:00 \\
 & & & & & 	1.9000 & 0.0700 & 3 & 98.21 & 2012-10-21T00:00:00 \\
 & & & & & 	1.7700 & 0.0800 & 3 & 109.74 & 2012-10-21T00:00:00 \\
 & & & & & 	2.0100 & 0.0300 & 3 & 98.21 & 2012-11-06T00:00:00 \\
 & & & & & 	1.9000 & 0.0400 & 3 & 109.74 & 2012-11-06T00:00:00 \\
 & & & & & 	1.9900 & 0.0300 & 3 & 98.21 & 2012-11-17T00:00:00 \\
 & & & & & 	1.8700 & 0.0400 & 3 & 109.74 & 2012-11-17T00:00:00 \\
 & & & & & 	1.9100 & 0.0500 & 3 & 109.74 & 2013-05-11T00:00:00 \\
 & & & & & 	1.9900 & 0.0300 & 3 & 98.21 & 2013-05-11T00:00:00 \\
 & & & & & 	2.1566 & 0.1078 & 3 & 91.8551 & 2013-05-31T09:07:08 \\
 ... & ... & ... & ... & ... & ... & ... & ... & ... & ... \\
\hline
\multicolumn{10}{l}{\textit{Notes}. $^{1}$ Classification: 1=Flat-spectrum radio quasar (FSRQ); 2=BL Lac; 3=BL Lac-galaxy dominated; 4=Blazar uncertain type;}\\
\multicolumn{10}{l}{5=BL Lac candidate; 6=Steep spectrum; 7=Uncertain.}\\
\multicolumn{10}{l}{$^{2}$ Observing date and time in the ISO standard format [YYYY-MM-DDThh:mm:ss], UTC time. Observations taken from the}\\
\multicolumn{10}{l}{``ALMA Calibrator Source Catalogue'', for which time information is not available, appear with time ``00:00:00''}\\
\end{tabular}
\label{tab:fluxes}
\end{table*}

In this paper we exploit the ACC catalogue for two purposes. In
Sect.~\ref{sect:validation} we use ALMA photometry to validate the new flux
density estimates presented in the \textit{Planck} multi-frequency Catalogue of
Non-Thermal sources \citep[PCNT;][]{PCNT}, and to assess the \textit{Planck}
completeness limits and  positional accuracy. So far, only an \textit{internal}
validation was possible and was indeed performed.

Next, we exploit the multifrequency ALMA measurements to estimate the
distribution of flux density ratios between Band 3 and higher frequency bands.
Such distributions allow us to extrapolate the observed 100\,GHz source counts
to higher frequencies where direct measurements are quite limited or missing
(Sect.~\ref{sect:nc}). Finally in Sect.~\ref{sect:conclusions} we present our
main conclusions.

\section{Validation of the \textit{Planck} Multi-frequency Catalogue of Non-thermal Sources}\label{sect:validation}

\begin{figure}
\begin{center}
\includegraphics[width=0.49\textwidth]{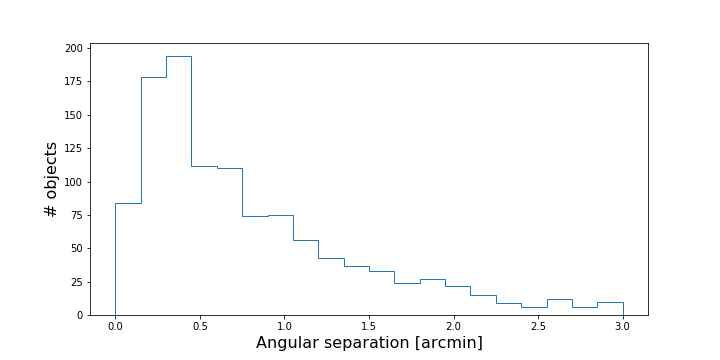}
\caption{Distribution of angular separations between ALMA and \citet{PCCS2} positions of the 1,120 sources resulting from the
cross-match ($3'$ search radius) between the ACC and the PCNT catalogues.}
 \label{fig:separation}
  \end{center}
\end{figure}

\begin{figure*}
\begin{center}
\includegraphics[width=0.49\textwidth]{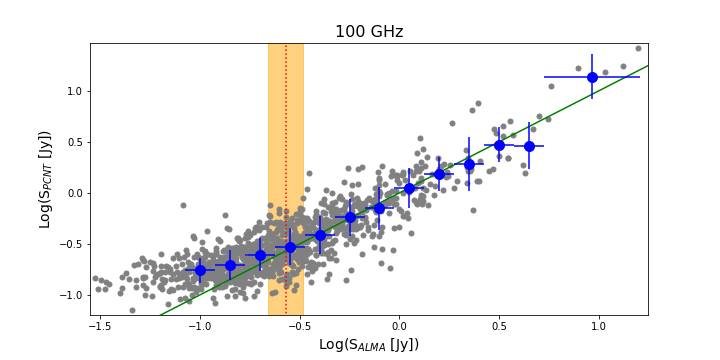}
\includegraphics[width=0.49\textwidth]{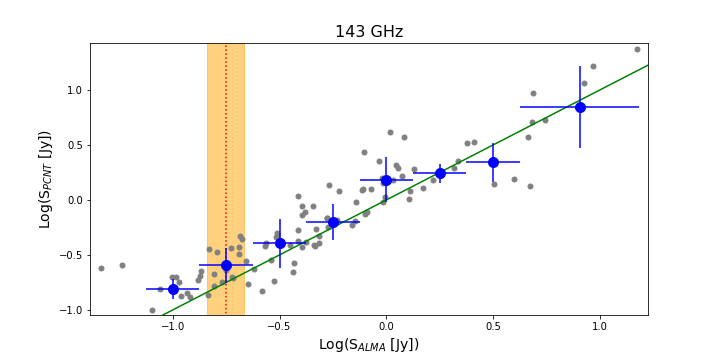}
\includegraphics[width=0.49\textwidth]{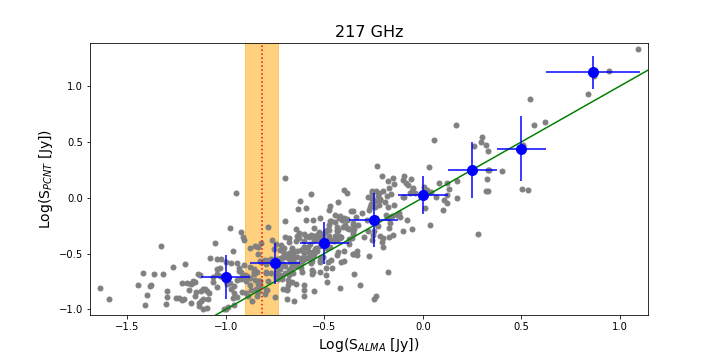}
\includegraphics[width=0.49\textwidth]{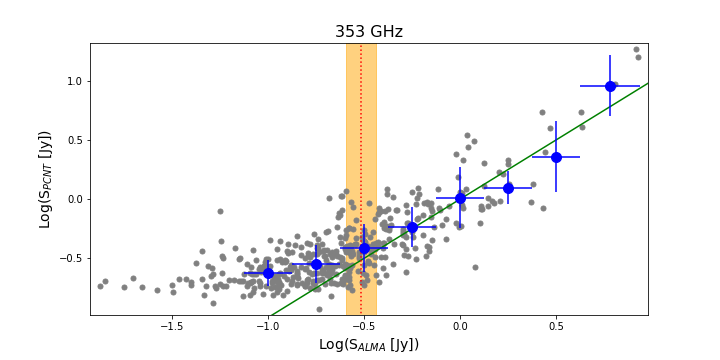}
\includegraphics[width=0.49\textwidth]{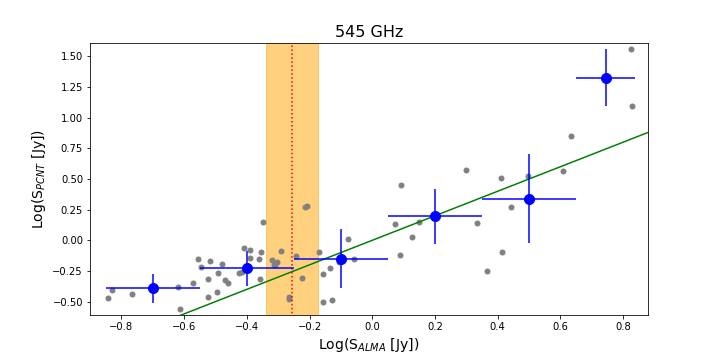}
\caption{Mean values of $\log(S_{\rm PCNT})$ (large blue circles with error bars) as a function of
$\log(S_{\rm ALMA})$ in bands 3, 4, 6, 7 and 8+9. The tiny
corrections for the slight differences between the effective frequencies of ALMA bands and those
of the nearest \textit{Planck} bands have been neglected except for Bands 7, 8 and 9.
Band 7 (285\,GHz) flux densities were extrapolated to 353\,GHz assuming the mean spectral
index ($-0.45; S\propto \nu^{-0.45}$) between band 6 and 8; Band 8 (467\,GHz) and 9 (673\,GHz) flux densities were extrapolated to 545\,GHz
using the mean spectral index ($-0.62$) between the two bands. The
frequency intervals are small so that the results are weakly dependent on the choice for the spectral index. The horizontal bars
correspond to the widths of $\log(S_{\rm ALMA})$ bins; the vertical bars show
the standard deviations around the mean values. The unbinned data points are
represented by small grey dots. The vertical dotted red lines
correspond to the PCCS2 90\% completeness limits with their uncertainties
represented by the shaded orange bands. The green solid lines correspond to
$S_{\rm PCNT}=S_{\rm ALMA}$. }
 \label{fig:validation}
  \end{center}
\end{figure*}

\begin{figure*}
\begin{center}
\includegraphics[width=0.49\textwidth]{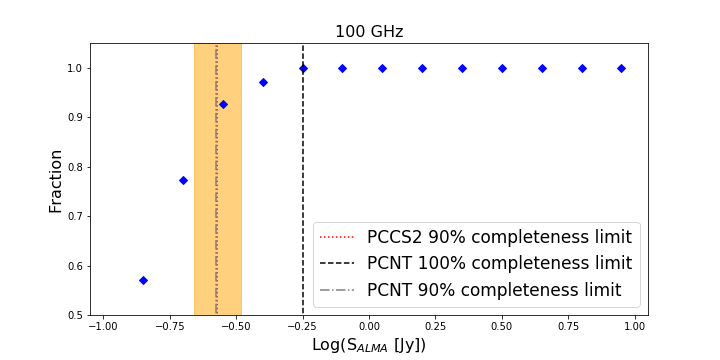}
\includegraphics[width=0.49\textwidth]{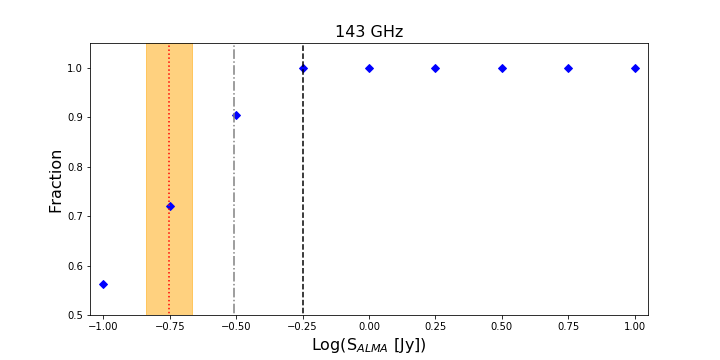}
\includegraphics[width=0.49\textwidth]{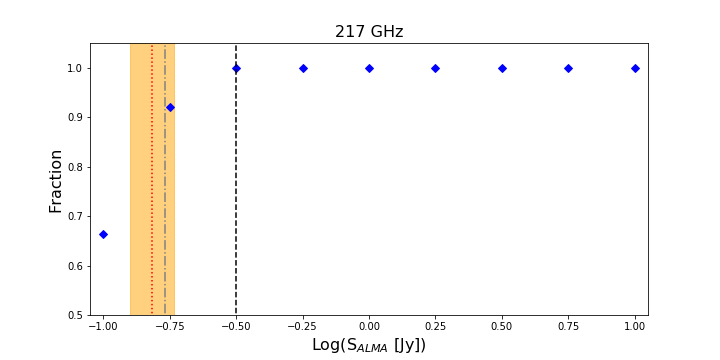}
\includegraphics[width=0.49\textwidth]{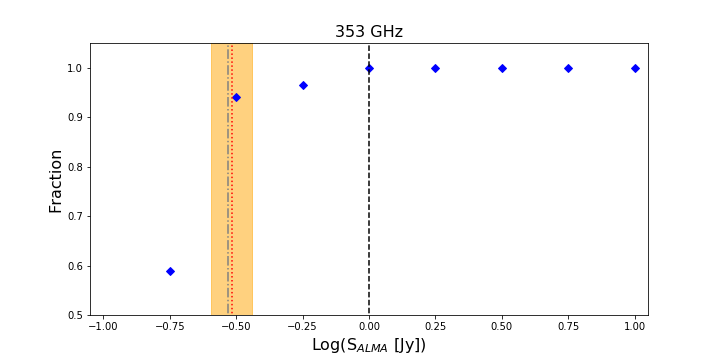}
\includegraphics[width=0.49\textwidth]{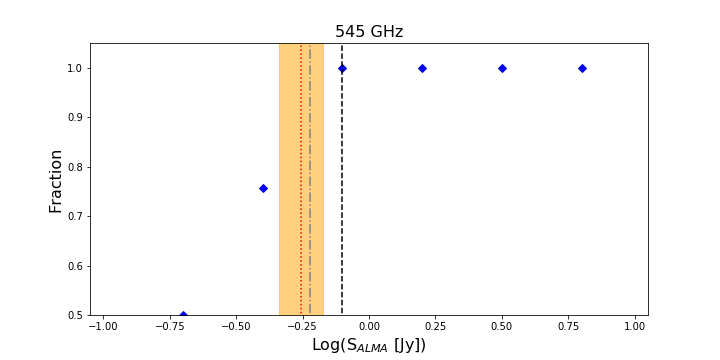}
\caption{Fraction of ACC sources with PCNT counterparts at 100, 143, 217, 353 and 545\,GHz as a function of ALMA flux density in Bands
3, 4, 6, 7 and $8+9$, respectively. The ALMA flux densities in the last two bands have been extrapolated
to 545\,GHz adopting the mean spectral index of our sources between the two bands, i.e. $S \propto \nu^{-0.62}$.
Band 7 flux densities have been extrapolated to 353\,GHz using the mean spectral index ($-0.45$) between Bands 6 and 8.
The vertical dashed black line and the dot-dashed grey line show, at each frequency, the flux densities
above which 100\% and 90\% of sources observed by ALMA have PCNT counterparts with $\hbox{SNR}\ge 3$. The dotted red lines show, for comparison,
the PCCS2 90\% completeness limits in the `extragalactic zone' with their uncertainties (shaded orange bands).
The number of sources observed by ALMA in band 10 is too small to allow meaningful estimates of the detection limits; see however the text.
}
 \label{fig:completeness}
  \end{center}
\end{figure*}

\begin{table*}
\centering
\caption{PCCS2 90\% completeness limits compared with the estimated PCNT 90\% and 100\% completeness limits,
defined as the ALMA flux densities above which 90\% or 100\%
of sources observed by ALMA have PCNT counterparts. Since there are too few sources
observed in the ALMA band 10 to derive meaningful limits, we give the 90\% and 100\%
limits derived by \citet{Maddox2018} from H-ATLAS data.}
\begin{tabular}{cccccc}
\hline
\hline
\textit{Planck} $\nu$ [GHz] & PCCS2 90\% &  PCNT 90\% & PCNT 100\% & H-ATLAS 90\%  & H-ATLAS 100\% \\
 & completeness &  completeness & completeness & completeness & completeness \\
 & [mJy] &  [mJy] & [mJy] & [mJy] & [mJy] \\
\hline
100	& 269 & 266 & 562 &  &   \\
143	& 177 & 312 & 562 &  &   \\
217	& 152 & 170 & 316 &  &   \\
353	& 304 & 296 & 1000 &  &   \\
545 & 555 &  597   &  1585   & &  \\
857	& 791 &     &     & 650 & 760 \\
\hline
\hline
\end{tabular}
\label{tab:completeness}
\end{table*}

The PCNT was built performing a multi-frequency analysis using the ``Matrix
Filters'' (MTFX) methodology (see \citealt{Herranz2008,Herranz2009}) at the
positions of sources detected by the Mexican Hat Wavelet 2 algorithm
\citep{GonzalezNuevo2006, LopezCaniego2006} in the full mission all-sky
temperature maps\footnote{The time frame of these data is $\sim$2.5\,years.}
with a signal-to-noise ratio (SNR) larger than 3 at either 30 and 143 GHz.  The
MTFX yielded flux densities and uncertainties in all nine \textit{Planck}
channels. This multifrequency approach has made it possible to reach
deeper detection limits, at given SNR, than the Second \textit{Planck}
Catalogue of Compact Sources \citep[PCCS2;][]{PCCS2} which contains sources
detected in each frequency channel separately.

So far the MTFX photometry could be validated only by comparison with flux
densities reported in the PCCS2, complemented with a statistical check made
comparing the number counts of catalogued sources with models. But PCCS2
estimates for PCNT sources are generally missing above 217\,GHz. Moreover the
amount of external data available to validate them was quite limited,
particularly at mm and sub-mm wavelengths.

To check the MTFX photometry up to high frequencies we cross-matched the ACC
with the PCNT outside the GAL070 mask, i.e. excluding the $\simeq 30\%$ of the
sky more heavily contaminated by Galactic emissions. We used a search radius of
$3'$, more than a factor of 3 larger than the estimated positional
uncertainties of \textit{Planck} sources \citep[$\lesssim 1'$, see tables 5 and
6 of][]{PCCS2}. As illustrated by Fig.\,\ref{fig:separation}, this positional
uncertainty estimate is confirmed by the comparison between ALMA and
\textit{Planck} positions. The positional differences peak at $\sim0'.38$ and
have a standard deviation $\sigma \simeq 0'.64$. The distribution is however
strongly asymmetric with an extended tail towards separations of a few arcmin.
This tail slightly decreases if sources below the PCCS2 90\% completeness
limits are excluded.

We found unique \textit{Planck} counterparts with $\hbox{SNR}\ge 3$ for 1,120
out of the 3,364 ALMA calibrators. Specifically, we found 1,069
counterparts (out of 3,310 calibrators) at 100\,GHz, 100 (out of 171) at
143\,GHz, 455 (out of 885) at 217\,GHz and 439 (out of 2,201) at 353\,GHz.
Moreover, 60 sources with ALMA measurements in band 8 or 9 have \textit{Planck}
counterparts at 545\,GHz. Obviously there is a large overlap among sources
observed in the different ALMA bands.

The mean surface density of PCNT sources outside the GAL070 mask is $\simeq
0.04\,\hbox{deg}^{-2}$, so that the probability that a PCNT source lies by
chance within $3'$ of an ALMA source is $\simeq 3\times 10^{-4}$. Thus the
expected number of spurious associations is $\simeq 1$, i.e. $\simeq 0.1\%$ of
\textit{Planck} counterparts.

\begin{table*}
\centering
\caption{Mean values of the log of the ratios of ALMA band 3 (characteristic frequency of 95\,GHz) flux densities to those measured in the other ALMA bands
and associated standard deviation, $\sigma$. By characteristic frequency of a band (2$^{\rm nd}$ column)
we mean the median frequency of the observations in such a band.}
\begin{tabular}{ccccc}
\hline
\hline
N. ALMA band & Charact. $\nu$ [GHz] & N. sources & $\langle\log({S_{\rm 95\,GHz}/S_{\nu}})\rangle$ & $\sigma$\\
\hline
4 & 145	& 164 & 0.11 & 0.10 \\
6 & 233	& 859 & 0.25 & 0.14  \\
7 & 285	& 2102 & 0.37 & 0.17 \\
8 & 467	& 87 & 0.35 & 0.18  \\
9 & 673	& 58 & 0.41 & 0.23 \\
\hline
\hline
\end{tabular}
\label{tab:ratio_nc}
\end{table*}

\begin{figure*}
\begin{center}
\includegraphics[width=0.49\textwidth]{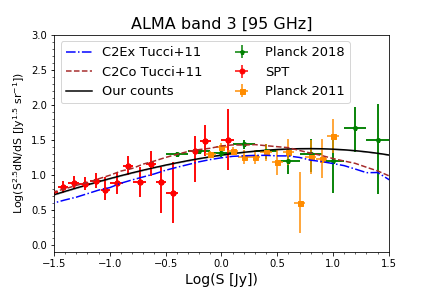}
\includegraphics[width=0.49\textwidth]{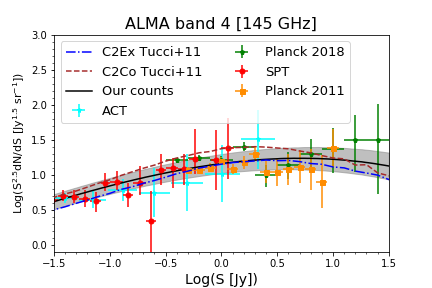}
\includegraphics[width=0.49\textwidth]{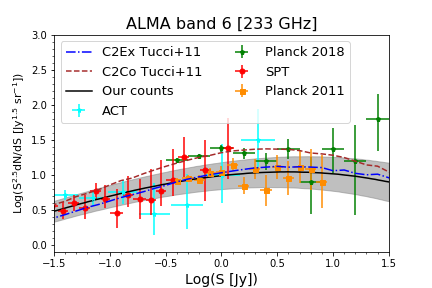}
\includegraphics[width=0.49\textwidth]{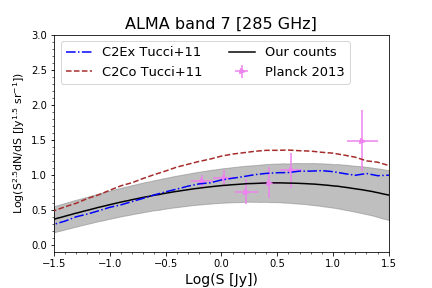}
\includegraphics[width=0.49\textwidth]{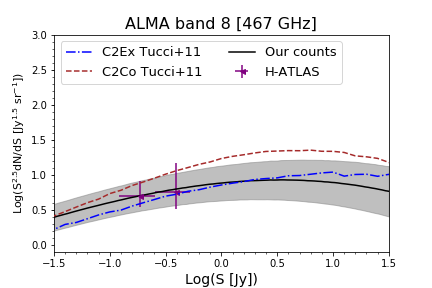}
\includegraphics[width=0.49\textwidth]{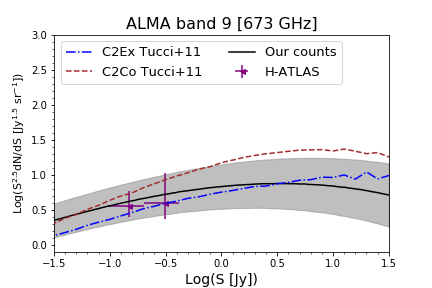}
\caption{Euclidean normalized differential source counts of radio sources in the ALMA bands for which measurements
for a substantial number of calibrators are available. The data are from \citet[][ACT]{Marsden2014},
\citet[][SPT]{Mocanu2013}, \citet{Planck2011}, \citet{Planck2013} and \citet{PCNT}. The band 8 and band 9 data points are our own estimates using Table~1 of
\citet{Negrello2017}; the extrapolations of the 600\,GHz ($500\,\mu$m) flux densities given
there to the effective frequencies of these 2 bands were done assuming
$S \propto \nu^{-0.62}$, i.e. using the mean spectral index between Bands 8 and 9. The mean spectral index between Bands 6 and 8 ($-0.45$)
was used to convert the \citet{Planck2013} counts from 353 to 285\,GHz.
The dot-dashed blue line and the dashed brown line show two models by \citet{Tucci2011}.}
 \label{fig:ratio1}
  \end{center}
\end{figure*}

MTFX flux density estimates at 100, 143, 217 and 353\,GHz were compared to ALMA
flux densities in Band 3, 4, 6 and 7, respectively. The small
differences of effective frequencies between Bands 3, 4 and 6 and the nearest
\textit{Planck} channels were neglected. Band 7 flux densities were converted
to 353\,GHz with $S\propto \nu^{-0.45}$, using the mean spectral index of our
sources between Bands 6 and 8. The \textit{Planck} 545\,GHz channel is
intermediate in frequency to ALMA Bands 8 and 9. We have extrapolated ALMA flux
densities in these bands to 545\,GHz adopting the mean spectral index of our
ALMA sources between these bands, i.e. with $S \propto
\nu^{-0.62}$. In any case, the extrapolations in frequency are quite small and the
spectral indices are relatively flat so that the results are only weakly
affected by chosen values.

For most sources we have several ALMA observations in each band. The comparison
with PCNT flux densities was made using the mean values and adopting the
standard deviation, summed in quadrature with the calibration uncertainty, as
an estimate of the uncertainty.


The results for \textit{Planck} channels from 100 to 545\,GHz are illustrated
by Fig.~\ref{fig:validation}. At the faintest flux density levels, the MTFX
photometry is affected by the Eddington bias \citep{Eddington1913,
HoggTurner1998}, which accounts for the increase of the \textit{Planck}/ALMA
flux density ratio with decreasing flux density. Above the PCCS2 90\%
completeness limits (vertical dotted red lines) we find good consistency
between the MTFX and the ALMA photometry. The large dispersion of flux density
ratios (typically $\sigma(\log(S_{\rm PCNT}/S_{\rm ALMA}))\simeq 0.2$) can be
entirely ascribed to variability\footnote{Blazars are known to be
strongly variable and \textit{Planck} and ALMA measurements are
non-simultaneous. \textit{Planck} flux densities are averages over the five
full sky surveys with the High Frequency Instrument (HFI) from 2009 August 12
to 2012 January 11. ALMA observations are distributed between  May 2011 and
July 2018.} plus measurement uncertainties, mostly on the \textit{Planck}
side. In particular, the few sources in the brightest luminosity bin have most
likely been caught by \textit{Planck} in a flaring
phase\footnote{Variability affects the \textit{Planck}/ALMA comparison
at all flux density levels. In well populated flux density bins its main effect
is to increase the dispersion of flux density ratios. The highest luminosity
bins, however, contain only a few sources whose mean flux density can be
dominated by extreme outbursts increasing flux densities by factors of ten or
more. This is the simplest explanation of anomalously high \textit{Planck}/ALMA
ratios such as the one in the brightest bin at 545\,GHz of
Fig.~\ref{fig:validation}. This bin contains only two sources, measured with
high signal-to-noise ratio both by \textit{Planck} and by ALMA, so that the
difference must be real.}.

The completeness of the PCNT catalog at each frequency $\geq 100\,$GHz was
tested by looking at the fraction of ACC objects having PCNT counterparts with
$\hbox{SNR}\ge 3$ as a function of the ALMA flux density. The results are shown
in Table\,\ref{tab:completeness} and in Fig.\,\ref{fig:completeness}. At 100,
217, 353 and 545\,GHz our results are in excellent agreement with the estimates
of the 90\% completeness limits given by \citet{PCCS2}.

At 143\,GHz the ALMA data suggest a significantly higher PCNT limit, but with
limited statistics (only 100 sources unevenly distributed among flux density
bins, to be compared with 1,069 sources at 100\,GHz, 455 at 217\,GHz and 439 at
353\,GHz). To assess the significance of the discrepancy we have performed
10,000 simulations of \textit{Planck} observations of the 100 ALMA calibrators.
These simulations were carried out randomly extracting each source from a
Gaussian distribution with mean value equal to its ALMA flux density and
dispersion equal to the mean PCNT uncertainty of those with a PCNT counterpart
($\sim$36\,mJy). A 90\% completeness limit within the uncertainty of the PCCS2
value was found in 10\% of the cases, implying that the discrepancy is only
marginally significant. We have also checked whether the difference of the
completeness limits may be due to the different photometric estimators used for
the PCCS2 and the PCNT (although the two photometries were shown to be, on
average, in good agreement by \citealt{PCNT}). To this end we have estimated
the PCCS2 completeness levels in the same way as we did for the PCNT. The
results did not change significantly.


At 857\,GHz, the poor statistics (only 4 calibrators with ALMA band 10 flux
density measurements have PCNT counterparts) hampers a reliable estimate.
However, an indication that the high frequency PCCS2 completeness limits are
conservative was provided by the cross-match of the \textit{Herschel}
Astrophysical Terahertz Large Area Survey (H-ATLAS) with the PCCS2 catalogue
\citep{Maddox2018}. These authors found PCCS2 90\% completeness limits of
$S_{\rm Herschel}=650\,$mJy at 857\,GHz ($350\,\mu$m). For comparison the PCCS2
estimate of the corresponding 90\% completeness limits is of 791\,mJy. Using
their catalogues, we further derived the 100\% PCCS2 completeness limits to be
of 760\,mJy at this frequency.


These results indicate that, although the MTFX approach reaches, at fixed SNR,
fainter flux densities than the single frequency approach adopted for the
PCCS2, and therefore the PCNT contains substantially more \textit {radio
sources}\footnote{The \textit{total} number of PCCS2 sources at high
frequencies is higher because it includes also dusty galaxies, while the PCNT
contains, by construction, only radio sources.}, the completeness limits remain
essentially unchanged.




\section{High frequency number counts of radio sources}\label{sect:nc}

The combination of \textit{Planck} \citep{PCNT} and SPT \citep{Mocanu2013} data
has provided an observational determination of radio source number counts at
$\simeq 100\,$GHz over a broad flux density range. An empirical description of
the Euclidean normalized counts at this frequency is (solid black line in the
top left panel of Fig.~\ref{fig:ratio1})
\begin{equation}\label{eq:100GHzcounts}
y=-0.01684\cdot x^{3}-0.1252\cdot x^{2}+0.2264\cdot x+1.282\, ,
\end{equation}
where $x=\log(S [{\rm Jy}])$ and $y=\log(S^{2.5}dN/dS [{\rm Jy}^{1.5}{\rm
sr}^{-1}])$.
We have exploited ALMA observations to extrapolate these counts to higher
frequencies where direct survey data are increasingly poor. To this end we have
calculated the mean values and the dispersions of the log of the flux density
ratios between band 3 (characteristic frequency of 95\,GHz) and the ALMA higher
frequency bands for which we have sufficient statistics. The results are shown
in Table~\ref{tab:ratio_nc}.

By convolving eq.~(\ref{eq:100GHzcounts}) with the appropriate distribution of
flux density ratios, assumed to be a Gaussian with mean and standard deviation given by
Table~\ref{tab:ratio_nc}, we have obtained the counts in bands 4, 6, 7, 8 and 9
shown by the solid black lines in Fig.~\ref{fig:ratio1}. The shaded grey bands
represent the $1\,\sigma$ uncertainties on the extrapolated
counts. These number counts are available in a
machine readable format in the website of  the Italian node of the European
ALMA Regional Centre (\url{http://arc.ia2.inaf.it}). Polynomial representations
of the extrapolated counts are given in Table\,\ref{tab:poly_parameters}.

The extrapolated counts are fully consistent with the available data. The
comparison with the two preferred models by \citet{Tucci2011}, ``C2Co'' and
``C2Ex'', shows that the latter performs quite well at all frequencies
while the former tends to overpredict the counts above 145 GHz.

\begin{table*}
\centering
\caption{Coefficients of the polynomial representations of the extrapolated
Euclidean normalized differential counts of radio sources at the effective frequencies of 5 ALMA bands:
$y=A\cdot x^{3}+B\cdot x^{2}+C\cdot x+D$, with $x=\log(S [{\rm Jy}])$ and $y=\log(S^{2.5}dN/dS [{\rm Jy}^{1.5}{\rm sr}^{-1}])$.}
\begin{tabular}{cccccc}
\hline
\hline
N. ALMA band & Charact. $\nu$ [GHz] & A & B & C & D \\
\hline
4 & 145	& -0.0166 & -0.128 & 0.207 & 1.160 \\
6 & 233	& -0.0160 & -0.133 & 0.175 & 0.988  \\
7 & 285	& -0.0159 & -0.136 & 0.151 & 0.846 \\
8 & 467	& -0.0159 & -0.135 & 0.159 & 0.883  \\
9 & 673	& -0.0157 & -0.134 & 0.156 & 0.833 \\
\hline
\hline
\end{tabular}
\label{tab:poly_parameters}
\end{table*}

\section{Conclusions}\label{sect:conclusions}

We have presented a new catalogue of ALMA flux density measurements of  radio
sources, mostly blazars, used as flux density, bandpass response, amplitude and
phase visibility calibrators. The catalogue was built combining the ALMACAL
catalogue published by \citet{Bonato2018} with the ALMA Calibrator Source
Catalogue. It contains ALMA observations for 3,364 bright, compact radio
sources  observed between May 2011 and July 2018, for a total of 47,115
observations in different bands and epochs. The catalogue is available online
as supplementary material and on the website of the Italian ALMA Regional
Center (ARC; \url{http://arc.ia2.inaf.it}).

We have added redshifts found in the literature, available for  2245 (67\%) of
the sources and a classification for all of them. The classification given in
the 5th edition of the Roma Multi-frequency Catalogue of Blazars
\citep[BZCAT;][]{Massaro2009} was adopted for the 1391 objects listed there. The
others were classified  following \citet{Bonato2018}.

The ALMA measurements were exploited to obtain the first \textit{external}
validation of the MTFX photometry presented in the new \textit{Planck}
multi-frequency Catalogue of Non-Thermal sources \citep[PCNT;][]{PCNT}, to
quantify its positional accuracy and to estimate its completeness limits.

We found good agreement between the ALMA and the MTFX photometry above the 90\%
completeness limits given by \citet{PCCS2}. The dispersions around the mean
MTFX/ALMA flux density ratios as a function of ALMA flux densities can be
accounted for by variability which also explains the excess flux densities
measured by \textit{Planck} for the few brightest sources, most likely detected
in a flaring phase. Below these limits, \textit{Planck} measurements show clear
signs of the Eddington bias.

The distribution of differences between ALMA and \textit{Planck} positions
peaks at $\sim0'.38$ and has a standard deviation $\sigma \simeq 0'.64$,
confirming the \citet{PCCS2} conclusion that the PCCS2 positional accuracy is
typically better than 1\,arcmin. The distribution has however an extended tail
reaching a few arcmin. The extension of such tail slightly decreases if sources
below the PCCS2 90\% completeness limits are excluded.

An analysis of the fraction of ALMA calibrators with a PCNT counterpart having
$\hbox{SNR}\ge 3$ at the nearest frequency as a function of the ALMA flux
density has shown that the PCNT completeness limits are consistent with the
PCCS2 ones at 100, 217, 353 and 545\,GHz. The PCNT limit at 143\,GHz seems to
be higher, but the difference is only marginally significant because of the
poor statistics.
We conclude that although the PCNT reaches fainter flux density levels than the
PCCS2, the completeness limits do not change appreciably.


Finally we have exploited the multi-frequency ALMA observations to derive the
distribution of flux density ratios between Band 3 and the higher frequency
bands. These distributions have allowed us to estimate the counts in such
bands, where direct measurements are limited or almost completely missing, by
extrapolating the relatively well determined 100\,GHz counts of radio sources.
The results agree with the available data and are consistent with the C2Ex
model by \citet{Tucci2011}, while their C2Co model is disfavoured.

\section*{Acknowledgements}

We thank the anonymous referee for a careful reading of the manuscript
and many constructive comments. This paper and the AKF and KAFE development
are part of the activities for the ALMA Re-Imaging Study approved in the
framework of the 2016 ESO Call for Development Studies for ALMA Upgrade (PI:
Massardi). The study acknowledges partial financial support by the Italian
Ministero dell'Istruzione, Universit\`{a} e Ricerca through  the  grant
`Progetti Premiali 2012 - iALMA' (CUP C52I13000140001). MB, MM and GDZ
acknowledge support from INAF under PRIN SKA/CTA FORECaST. GDZ acknowledges
support from ASI/INAF agreement n.~2014-024-R.1 for the {\it Planck} LFI
Activity of Phase E2 and from the ASI/Physics Department of the university of
Roma--Tor Vergata agreement n. 2016-24-H.0 for study activities of the Italian
cosmology community. MN acknowledges support from the European Union's Horizon
2020 research and innovation programme under the Marie Sk{\l}odowska-Curie
grant agreement No 707601. DH thanks the Spanish MINECO for partial financial
support under project AYA2015-64508-P and funding from the European Union’s
Horizon 2020 research and innovation programme (COMPET-05-2015) under grant
agreement number 687312 (RADIOFOREGROUNDS). LB and JGN acknowledge financial
support from the I+D 2015 project AYA2015-65887-P (MINECO/FEDER). JGN also
acknowledges financial support from the Spanish MINECO for a ‘Ramon y Cajal’
fellowship (RYC-2013-13256). This paper makes use of the following ALMA data: ADS/JAO.ALMA\#2011.0.00001.CAL. ALMA is a partnership of ESO (representing its member states), NSF (USA) and NINS (Japan), together with NRC (Canada), MOST and ASIAA (Taiwan), and KASI (Republic of Korea), in cooperation with the Republic of Chile. The Joint ALMA Observatory is operated by ESO, AUI/NRAO and NAOJ.

\bibliographystyle{mnras}
\bibliography{biblio} 

\bsp	
\label{lastpage}
\end{document}